\documentclass[10pt,twocolumn,aps,prb,showpacs,groupedaddress]{revtex4-1}
\usepackage{epsfig}
\usepackage{float}
\usepackage{color}
\usepackage{multirow}
 
\begin{document}

\title{Single crystal investigation of proposed type-II Weyl semimetal CeAlGe}

\author{H. Hodovanets,$^{1}$ C. J. Eckberg,$^{1}$ P. Y. Zavalij,$^{1,2}$ H. Kim,$^{1}$ W.-C.~Lin,$^{1}$ M. Zic,$^{1}$ D. J. Campbell,$^{1}$ J. S. Higgins,$^{1}$ and J. Paglione$^{1}$}

\affiliation{$^1$ Center for Nanophysics and Advanced Materials, Department of Physics, University of Maryland, College Park 20742, USA}
\affiliation{$^2$ X-ray Crystallographic Center, Department of Chemistry and Biochemistry, University of Maryland, College Park 20742, USA}

\begin{abstract}

We present details of materials synthesis, crystal structure, and anisotropic magnetic properties of single crystals of CeAlGe, a proposed type-II Weyl semimetal. Single-crystal x-ray diffraction confirms that CeAlGe forms in noncentrosymmetric I4$_1$md space group, in line with predictions of non-trivial topology. Magnetization, specific heat and electrical transport measurements were used to confirm antiferromagnetic order below 5 K, with an estimated magnon excitation gap of $\Delta$~=~9.11~K from heat capacity and hole-like carrier density of 1.44~$\times$~10$^{20}$~cm$^{-3}$ from Hall effect measurements. The easy magnetic axis is along the [100] crystallographic direction, indicating that the moment lies in the tetragonal \textit{ab}-plane below 7~K. A spin-flop transition to less than 1~$\mu_B$/Ce is observed to occur below 30~kOe at 1.8 K in the $M(H)$ (\textbf{H}$\|$\textbf{a}) data. Small magnetic fields of 3~kOe  and 30~kOe are sufficient to suppress magnetic order when applied along the \textit{a}- and \textit{c}-axes, respectively, resulting in a complex \textit{T-H} phase diagram for \textbf{H}$\|$\textbf{a} and a simpler one for \textbf{H}$\|$\textbf{c}. 

\end{abstract}


\maketitle

\section{Introduction}

The term Weyl fermions refers to a pair of relativistic fermions with different chiralities or handness. Although these conceptual fermions were predicted in a framework of high energy physics, Weyl fermions are believed to be realized in the solid state, i.e., Weyl semimetals type-I and type-II.\cite{Hasan2017,Chang2017,Yan2017} In these two types, broken symmetry of either parity or time-reversal reduces Dirac fermions to Weyl fermions. There are some cases where both symmetries are simultaneously broken.\cite{Liu2014a,Neupane2014} Although application of magnetic field breaks time-reversal symmetry in an inversion-symmetry broken Weyl semimetal, in the case of CeAlGe and PrAlGe, both symmetries break due to noncentrosymmetric crystal structure and intrinsic magnetic order of the \textit{f} magnetic moment.\cite{Chang2018} Thus, CeAlGe and PrAlGe provide a new route to generating type-II Weyl fermions. 

CeAlGe was theoretically predicted to order ferromagnetically with the easy axis along the crystallographic \textit{a}- axis.\cite{Chang2018} To the best of our knowledge only limited measurements on polycrystalline samples of CeAlGe are reported. CeAlGe was found to order antiferromagnetically (AFM) below 5~K and form in a tetragonal structure of I4$_1$md (noncentrosymmetric)\cite{Zhao1990,Dhar1992,Dhar1996} space group and, more recently, to be a soft ferromagnet with Curie temperature of 5.6~K with the tetragonal unit cell of I4$_1$/amd (centrosymmetric) space group\cite{Flandorfer1998} [here CeAlGe is a variant of binary substitution Ce(Al$_{0.5}$Ge$_{0.5}$)$_2$].  While the former structure provides a potential route for the realization of nontrivial topology in this system, the latter structure does not. Thus, it is important to address the issue of the space group in this system especially since the easily accessible and widely used powder x-ray diffraction measurements are not able to distinguish between the two. In order to address the nature of the magnetic order, determine the intrinsic crystal structure, and study anisotropic properties of CeAlGe, single crystals were grown. Single crystal x-ray measurements were performed, showing that CeAlGe forms in the tetragonal crystal structure of I4$_1$md, noncentrosymmetric, space group, confirming the possibility of type-II Weyl fermions in this system. According to our measurements, CeAlGe orders antiferromagnetically below 5~K, and the \textit{a}-axis is the easy axis indicating that the moment lies in the tetragonal \textit{ab}-plane below 7~K and 50~kOe. We observe a spin-flop transition for \textbf{H}$\|$\textbf{a} in the $M(H)$ data with the saturation moment reaching about 0.8~$\mu_B$ at 140~kOe and 1.8~K (for comparison, saturated moment for \textbf{H}$\|$\textbf{c} is about 1.2~$\mu_B$). We report detailed temperature- and field-dependent magnetization, resistivity, Hall effect, and heat capacity measurements which allowed us to construct tentative \textit{T-H} phase diagrams for \textbf{H}$\|$\textbf{a} and \textbf{H}$\|$\textbf{c}.
 
%----------------------------------------------------------------------------------------------------------
%----------------------------------------------------------------------------------------------------------
\section{Experimental details}

\begin{figure*}[th]
\centering
\includegraphics[width=1\linewidth]{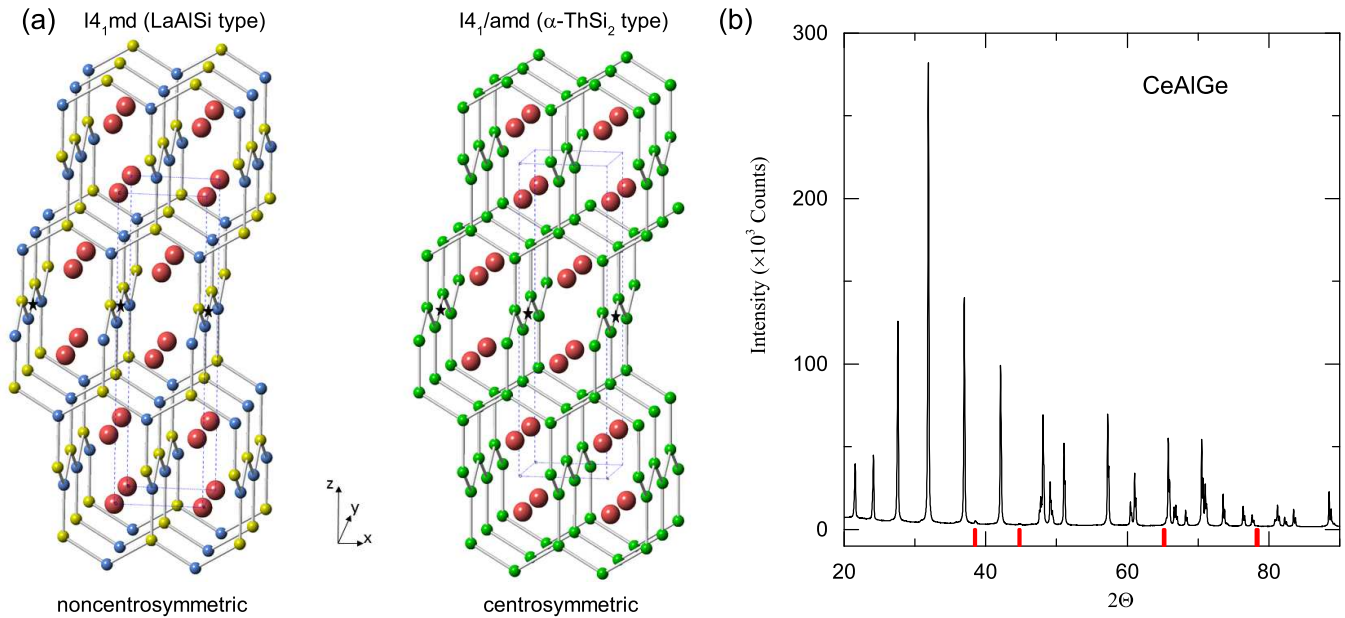}
\caption{\footnotesize (color online) (a) Schematic presentation of nonsentrosymmetric, I4$_1$md, LaAlSi type, and centrosymmetric I4$_1$/amd, $\alpha$-ThSi$_2$ type, space groups of CeAlGe. Ordered structure in I4$_1$md with Ge (blue) Al (yellow) is shown on the left. Disordered structure in I4$_1$/amd with Al and Ge mixed 50:50 in each position (green balls) is and shown on the right. Respective unit cells are shown with dashed lines. The center of symmetry (inversion center, shown by black stars), if placed in the middle of Ge-Al bond of I4$_1$/md space group, will transform blue Ge into yellow Al giving green mixture Ge/Al and centrosymmetric I4$_1$/amd space group. (b) Powder x-ray diffraction pattern of ground CeAlGe single crystals. The few low-intensity peaks marked with lines are due to the Al flux.}
\label{XRD}
\end{figure*}

Single crystals were grown by the high temperature Al self-flux method.\cite{Canfield1992,Canfield2010,Canfield2001} Chunks of Ce/La (99.8$\%$, AlfaAesar), shots of Ge (99.999$\%$ metals basis, AlfaAesar), and pieces of Al (99.999$\%$ metal basis, AlfaAesar) in the ratio of 10:10:80 were placed in the Canfield alumina crucible set with a decanting frit (LSP ceramics), sealed in the quartz ampule under partial Ar pressure, heated to 1150~$^o$C, held at that temperature for 2~h, cooled down at 5~deg/h to 750~$^o$C with subsequent decanting of the excess Al with the help of a centrifuge. Note that samples grown using quartz wool as a strainer (rather than a ceramic frit) show different magnetic properties, suspected to be a result of small Si substitution. Therefore, all measurements presented in this work are for single crystals grown using quartz wool-free Canfield crucible sets with the frits. LaAlGe single crystals were grown as a non-\textit{f} local moment bearing analog. Single crystals grow as large plates with a mirror-like surface and the \textit{c}-axis perpendicular to the plates and naturally formed edges of the plates being tetragonal \textit{a} and \textit{b} axes. All measurements presented for LaAlGe and CeAlGe were done on the samples from the same batch, respectively. 

A Rigaku MiniFlex diffractometer (Cu radiation) with DeTEX detector was used to collect powder x-ray diffraction patterns to confirm single phase, apart from the Al flux, present. Single crystal x-ray intensity data were measured at 250~K on a Bruker APEX-II CCD system equipped with a graphite monochromator and a MoK$\alpha$ sealed tube (wavelength $\lambda$ = 0.71070~$\mathrm{\AA}$). The structure was solved and refined using the Bruker SHELXTL software package.

Temperature- and field-dependent resistivity, Hall effect, magnetization, and specific heat measurements were performed in the 300 to 1.8~K temperature range and magnetic field up to 140~kOe in a commercial cryostat. Resistivity measurements were made in a standard four-probe geometry ($I$ = 1~mA). For resistivity and Hall effect measurements, the samples were polished and shaped with care to not have any Al inclusions. Crystals are rather brittle and extreme caution is needed when polishing. Electrical contact to the samples was made with Au wires attached to the samples using EPOTEK silver epoxy and subsequent cure at 100~$^0$C. Electrical current was directed along a [010] direction of the tetragonal \textit{ab}-plane, naturally formed edge of the plates. We used the cross-sectional area of the samples and the distance between the midpoint of two voltage contacts to calculate the resistivity of the samples. 

The magnetization was measured using a vibrating sample magnetometer (VSM). The samples were mounted using GE varnish. The contribution of the GE varnish to the $M(T)/H$ and $M(H)$ data was assumed to be negligible. 

The dynamic susceptibility was measured using AC option of Quantum Design Magnetic Properties Measurement System (MPMS$\textsuperscript{\textregistered}$3).

\begin{table*}
\caption{\label{tabl1} CeAlGe and LaAlGe crystallographic data determined through single-crystal x-ray diffraction. All data were collected at 250~K on Bruker APEX-II CCD system equipped with a graphite monochromator and a MoK$\alpha$ sealed tube (wavelength $\lambda$~=~0.71070~$\mathrm{\AA}$). The refinement results are given for CeAlGe in two different reported structures for comparison.}
\footnotesize\rm
\begin{ruledtabular}
\begin{tabular}{llll}
Chemical formula&CeAlGe&CeAl$_{1.05}$Ge$_{0.95}$&LaAlGe\\
Space group&I4$_1$md (No. 109)&I4$_1$/amd (No. 141)&I4$_1$md (No. 109)\\
Structure type&LaPtSi&$\alpha$-ThSi$_2$&LaPtSi\\
\hline
$a$($\mathrm{\AA}$)&4.2920(2)&4.2920(2)&4.3337(2)\\
$b$($\mathrm{\AA}$)&4.2920(2)&4.2920(2)&4.3337(2)\\
$c$($\mathrm{\AA}$)&14.7496(4)&14.7496(4)&14.8097(6) \\
$V^3$($\mathrm{\AA}^3)$&271.71(3)&271.71(3)&278.14(3) \\
Z&4&4&4\\
Reflections collected&3197 [R$_{int}$=0.0202]&3349 [R$_{int}$=0.0344]&3415 [R$_{int}$=0.0320]\\
Data/restraints/parameters&380/1/12&207/0/9&404/1/15\\
Goodness-of-fit on F$^2$&1.260&1.326&1.000 \\
Final R indexes [I$>$=2$\sigma$(I)]&R$_1$=0.0175, wR$_2$=0.0457 &R$_1$=0.0245, wR$_2$=0.0648& R$_1$=0.0178, wR$_2$=0.0419 \\
Final R indexes [all data]& R$_1$=0.0207, wR$_2$=0.0486 &R$_1$=0.0258, wR$_2$=0.0668&R$_1$=0.0209, wR$_2$=0.0437 \\
Largest diff. peak/hole / e${\mathrm{\AA}^{-3}}$ &1.06/-2.04& 2.822/-3.452&1.51/-1.35 \\
\textit{Flack parameter}&0.02(4)&&0.01(4)\\\
\end{tabular}
\end{ruledtabular}
\end{table*}

\begin{table}
\caption{\label{tabl2} Fractional Atomic Coordinates and Equivalent Isotropic Displacement Parameters ($\AA^2$) for CeAlGe and LaAlGe. U$_{eq}$ is defined as 1/3 of of the trace of the orthogonalised U$_{IJ}$ tensor.}
\footnotesize\rm
\begin{ruledtabular}
\begin{tabular}{llllll}
\multicolumn{6}{c}{Space group I4$_1$md}\\
\multicolumn{6}{c}{Chemical formula CeAlGe}\\
Atom&\textit{x}&\textit{y}&\textit{z}&{U(eq)}&Occ.\\
Ce&0.5&0.5&0.49969(11)&0.00598(17)&1.0\\
Al&0&0&0.5829(4)&0.00879(18)&1.0\\
Ge&0&0&0.41697(2)&0.00879(18)&1.0\\
\hline
\multicolumn{6}{c}{Space group I4$_1$/amd}\\
\multicolumn{6}{c}{Chemical formula CeAl$_{1.05}$Ge$_{0.95}$}\\
Atom&\textit{x}&\textit{y}&\textit{z}&{U(eq)}&Occ.\\
Ce&0.5&0.75&0.375&0.0070(3)&1.0\\
Al&0.5&0.25&0.20773(7)&0.0093(5)&0.522(9)\\
Ge&0.5&0.25&0.20773(7)&0.0093(5)&0.478(9)\\
\hline\hline
\multicolumn{6}{c}{Space group I4$_1$md}\\
\multicolumn{6}{c}{Chemical formula LaAlGe}\\
Atom&\textit{x}&\textit{y}&\textit{z}&{U(eq)}&Occ.\\
La&0.5&0.5&0.49999(12)&0.00738(15)&1.0\\
Al&0&0&0.4168(4)&0.0128(16)&1.0\\
Ge&0&0&0.58304(2)&0.0103(6)&1.0\\\
\end{tabular}
\end{ruledtabular}
\end{table}

%----------------------------------------------------------------------------------------------------------
%----------------------------------------------------------------------------------------------------------

\section{Results}
 
\begin{figure}[th]
\centering
\includegraphics[width=1\linewidth]{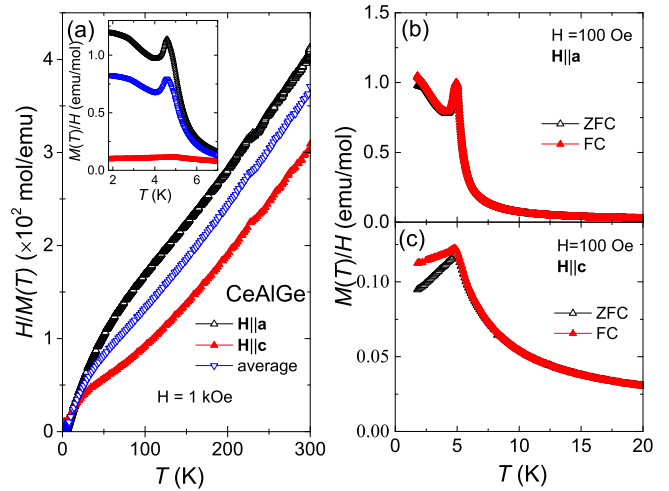}
\caption{\footnotesize (color online) (a) Temperature-dependent inverse magnetization of CeAlGe for \textbf{H}$\|$\textbf{a} and \textbf{H}$\|$\textbf{c}, together with the data for the polycrystalline average. The inset shows low-temperature part of the $M(T)/H$ data. (b) and (c) Zero-field cooled (ZFC) and field-cooled (FC) data for \textbf{H}$\|$\textbf{a} and \textbf{H}$\|$\textbf{c}, respectively.}
\label{MH}
\end{figure}

\begin{figure*}[th]
\centering
\includegraphics[width=1\linewidth]{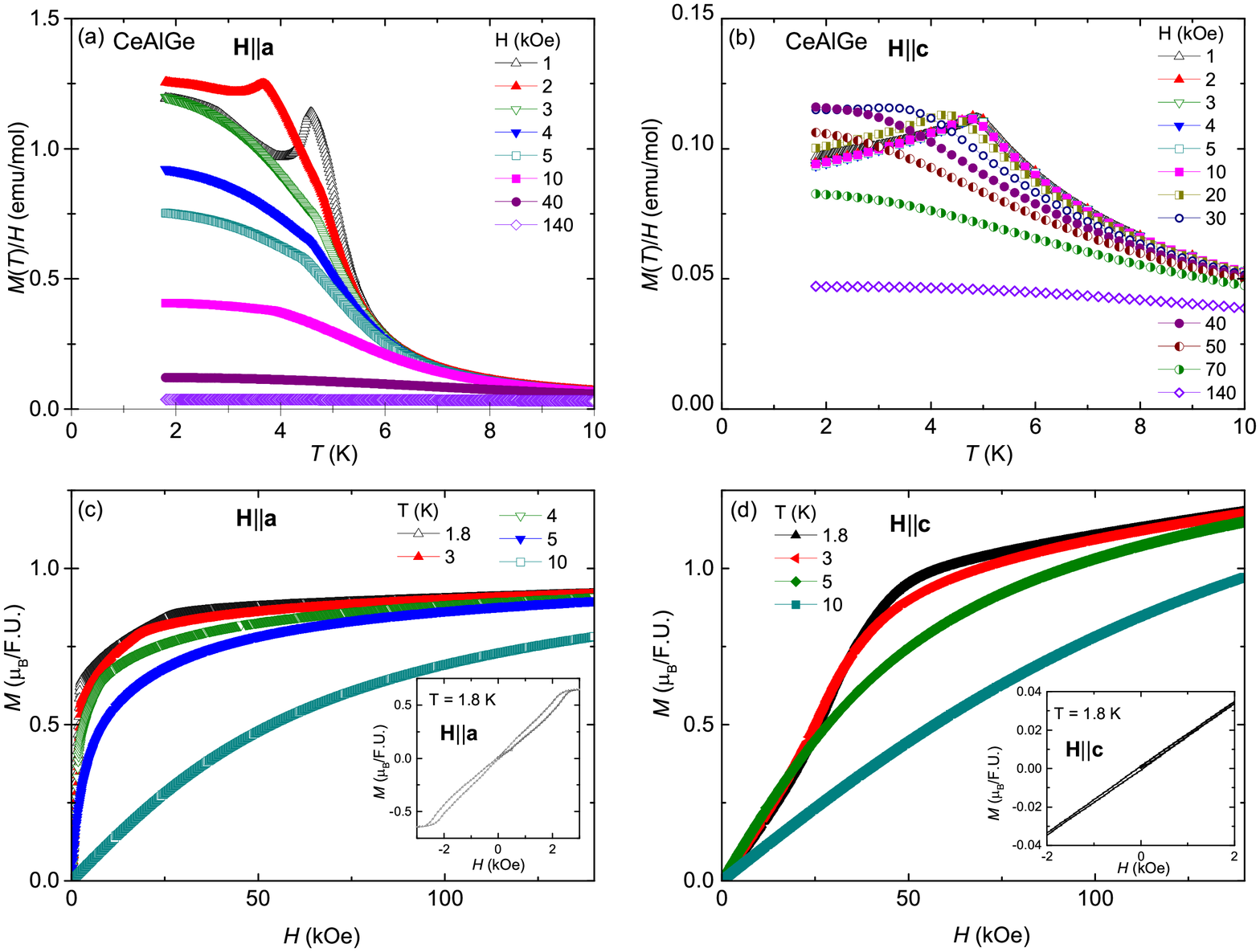}
\caption{\footnotesize (color online) (a) and (b) temperature-dependent magnetization of CeAlGe at constant magnetic fields for \textbf{H}$\|$\textbf{a} and \textbf{H}$\|$\textbf{c}, respectively. (c) and (d) Field-dependent magnetization of CeAlGe single crystal for \textbf{H}$\|\textbf{a}$ and \textbf{H}$\|$\textbf{c}, respectively.}
\label{MH1}
\end{figure*}

\subsection{Crystal structure}
A schematic comparison of noncentrosymmetric, I4$_1$md, LaAlSi type, and centrosymmetric I4$_1$/amd, $\alpha$-ThSi$_2$ type,  space groups is shown in Fig.~\ref{XRD}(a) with respective unit cells shown in dashed lines. For the ordered structure, I4$_1$md, Ge and Al atoms are shown in blue and yellow, respectively, and for the disordered structure, I4$_1$/amd, Al and Ge are mixed 50:50 in each position and are shown in green. Apart from the color of the Ge and Al atoms, if one looks at the extended crystal structure, they are the same. The center of symmetry (inversion center), which if placed in the middle of Ge-Al bond (as shown by black stars) will transform blue Ge into yellow Al giving green mixture Ge/Al and centrosymmetric group. As a consequence, the powder x-ray pattern for both crystal structures is identical save for the slight difference in intensities for some of the reflections and one cannot determine which crystal structure it is based on the powder x-ray pattern alone. Thus, in Fig.~\ref{XRD}(b), we show the powder diffraction pattern for CeAlGe for the purpose of confirming that there are no other secondary phases, contribution of Al flux excluded. In order to be able to resolve between the two space groups, we performed single crystal x-ray diffraction. The results of the CeAlGe and non-\textit{f} moment bearing LaAlGe single crystal x-ray refinement are given in Table~\ref{tabl1} and the atomic position for all atoms are given in Table~\ref{tabl2}. First, we would like to address the crystal structure. Final R indexes and goodness of fit on $R^2$ are small numbers which indicate that both crystal structures are likely. However, to distinguish between the two and confirm noncentrosymmetric structure, we looked at the Flack parameter\cite{Flack1983} that allows for an estimation of the absolute configuration of a structural model and is only defined for noncentrosymmetric unit cell. If the value of the Flack parameter is near 0 with small uncertainty, which is the case here, the absolute structure determined by single crystal structure refinement is likely correct, indicating that CeAlGe forms in noncentrosymmetric crystal structure, I4$_1$md. 
The lattice parameters determined for LaAlGe are consistent with the ones reported in the literature.\cite{Guloy1991} The lattice parameters of CeAlGe determined here are larger than the ones found in the literature\cite{Flandorfer1998,Dhar1996} for polycrystalline samples. We believe that Si doping from quartz wool used for the growth may be a culprit that led to the smaller reported lattice parameters of the polycrystalline samples (the lattice parameters of CeAlSi are smaller\cite{Bobev2005} than those of CeAlGe). 

%----------------------------------------------------------------------------------------------------------
%----------------------------------------------------------------------------------------------------------
\subsection{Magnetization}

Figure~\ref{MH}(a) shows the temperature-dependent $H/M(T)$ data for \textbf{H}$\|$\textbf{a}, \textbf{H}$\|$\textbf{c} and a polycrystalline average calculated as (2$\chi_a$+$\chi_c$)/3. The N\'{e}el temperature, $T_N$, is seen as a maximum in the $M(T)/H$ data shown in the inset to Fig.~\ref{MH}(a). The temperature-dependent moment is highly anisotropic, with the $M(T)/H$ (\textbf{H}$\|$\textbf{a}) being about 10 time larger than that of $M(T)/H$ (\textbf{H}$\|$\textbf{c}) at the magnetic ordering temperature. The effective moment estimated from the Curie-Weiss fit of the $M(T)/H$ data of the polycrystalline average above 150~K results in $\mu_{eff}$ = 2.56~$\mu_B$, indicating Ce$^{3+}$, and paramagnetic Weiss temperature of $-$3.6~K, indicating dominant antiferromagnetic interactions.  

Figures~\ref{MH}(b) and (c) show zero-field cooled (ZFC) and field-cooled (FC) $M(T)/H$ data at 100~Oe for \textbf{H}$\|$\textbf{a} and \textbf{H}$\|$\textbf{c}, respectively. For the $M(T)/H$ (\textbf{H}$\|$\textbf{a}) data, there is a very small hysteresis below $\sim$~2.5~K. As for the $M(T)/H$ (\textbf{H}$\|$\textbf{c}) data, ZFC and FC data show different behavior in the ordered state, which may be due to the domain reorientation.

\begin{figure}[tb]
\centering
\includegraphics[width=1\linewidth]{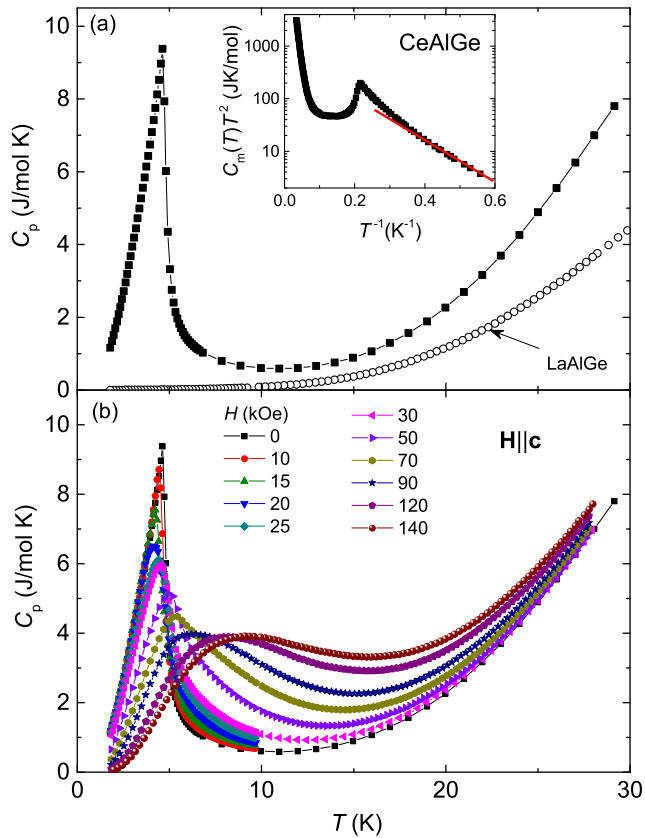}
\caption{\footnotesize (color online) (a) Temperature-dependent heat capacity data of CeAlGe and LaAlGe single crystals. The inset shows $C_m(T)T^2$ vs $T^{-1}$ data for CeAlGe on a semi-log scale. A line drawn through the low-temperature data points shows the low-temperature fit for the estimate of the phenomenological gap $\Delta$ in the magnon excitation spectrum. The fit is based on the phenomenological form of $C_m(T)\approx (1/T^2)e^{-\Delta/T}$. $\Delta$ was found to be 9.11~$\pm$~0.09~K. (b) Temperature-dependent heat capacity of CeAlGe single crystal at constant magnetic fields for \textbf{H}$\|$\textbf{c}. }
\label{C}
\end{figure}

\begin{figure*}[th]
\centering
\includegraphics[width=1\linewidth]{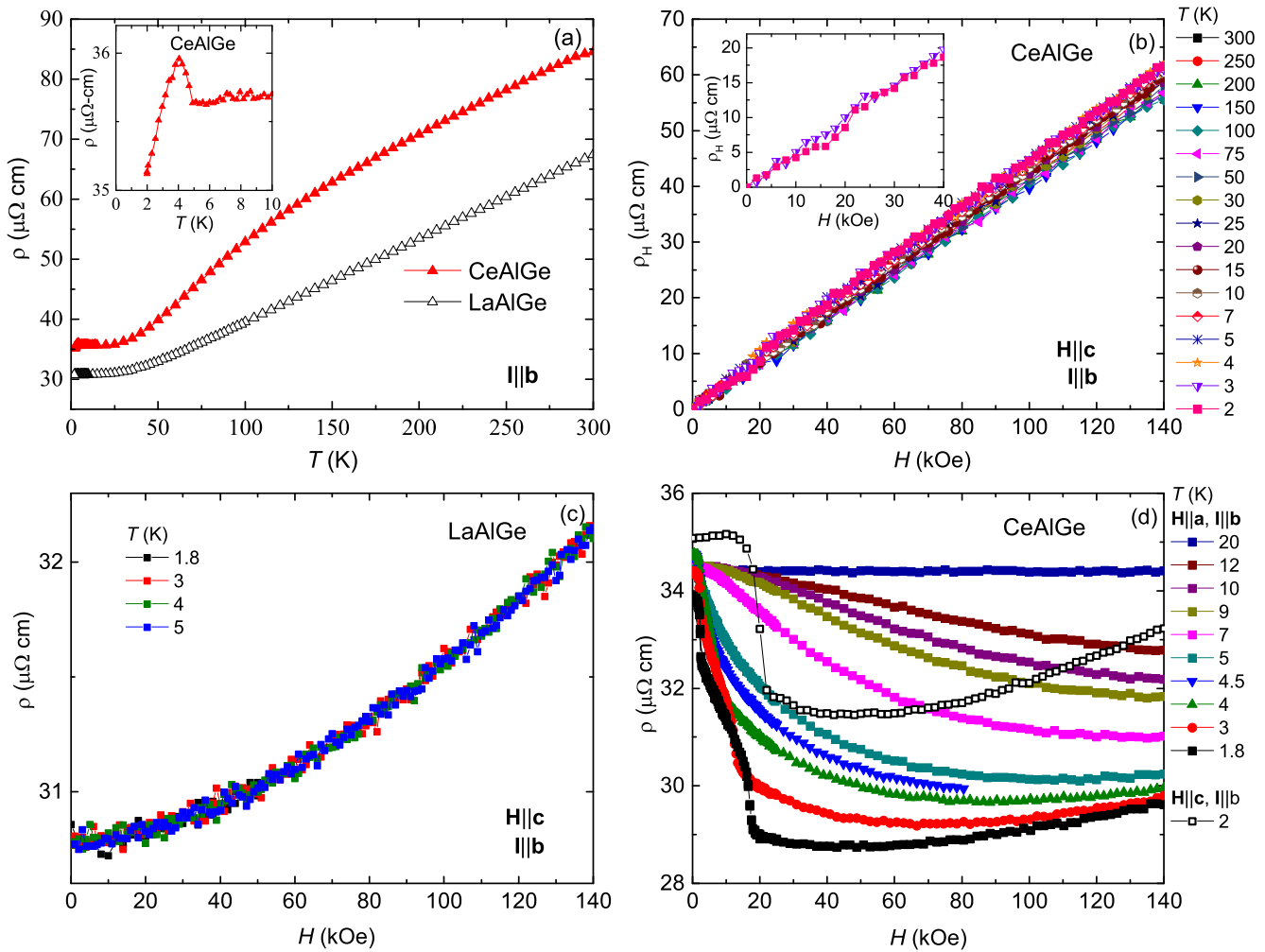}
\caption{\footnotesize (color online) (a) Temperature-dependent resistivity of CeAlGe and LaAlGe single crystals. The inset shows the feature associated with the magnetic order below 5~K in the resistivity data. (b) Field-dependent Hall resistivity of CeAlGe single crystal at constant temperatures for \textbf{H}$\|$\textbf{c}. Inset shows the Hall resistivity at 2 and 3~K, with the features just below 20~kOe, which are associated with the suppression of the magnetic order. (c) Transverse magnetoresistivity of LaAlGe single crystal at constant temperatures, \textbf{H}$\|$\textbf{c}. (d) Transverse magnetoresistivity of CeAlGe single crystal at constant temperatures for \textbf{H}$\|$\textbf{a} and \textbf{H}$\|$\textbf{c} (only the data collected at 2~K).}
\label{R}
\end{figure*}

The features associated with the magnetic order are suppressed by 3~kOe and 40~kOe for magnetic field applied along the \textit{a} and \textit{c}-axes, Figs.~\ref{MH1}(a) and (b), respectively. The effect of the field is more drastic for \textbf{H}$\|$\textbf{a}, where, as the magnetic field is increased beyond 3 kOe, the $M(T)/H$ data saturate at low temperature and the magnitude is substantially reduced. At the highest field of 140~kOe, the $M(T)/H$ data for both directions are nearly isotropic.

Field-dependent magnetization $M(H)$ data at constant temperatures for the field along the \textit{a} and \textit{c} axes are shown in Figs.~\ref{MH1}(c) and (d), respectively. At the lowest temperature of 1.8~K, for \textbf{H}$\|$\textbf{a}, a clear sharp spin-flop transition to a less than 1~$\mu_B$ saturation moment is observed below 2.5~kOe. After this transition, the moment remains almost constant up to 140~kOe, indicating that higher magnitude field is necessary to observe one or more metamagnetic transitions to reach a value of about 2.14~$\mu_B$ expected for the free-ion Ce$^{3+}$ full saturated moment. As the temperature is increased, the region of the canted phase of the spin-flop transition becomes smaller and is almost indistinguishable at 5~K. The inset to Fig.~\ref{MH1}(c) shows a hysteresis curve, indicating that below 2.5~kOe, the transition is of the first order. This also explains the different behavior in the ZFC and FC data below 2.5~K in Fig.~\ref{MH}(b). The metamagnetic-like transition, associated with the suppression of the magnetic order is also observed in the $M(H)$ data for \textbf{H}$\|$\textbf{c}, Fig.~\ref{MH1}(d). As in the case of \textit{M(H)} data for \textbf{H}$\|$\textbf{a}, the magnetic moment in the \textit{M(H)} data for \textbf{H}$\|$\textbf{c} reaches a little over 1~$\mu_B$ by 140~kOe. There appear to be a very small hysteresis in the \textit{M(H)} curve, \textbf{H}$\|$\textbf{c}, shown in the inset to Fig.~\ref{MH1}(d), which accounts for ZFC and FC data shown in Fig.~\ref{MH}(c).

%----------------------------------------------------------------------------------------------------------
%----------------------------------------------------------------------------------------------------------
\subsection{Heat capacity}

The temperature-dependent heat capacity data for CeAlGe and non-local moment bearing LaAlGe single crystals are shown in Fig.~\ref{C}(a). The linear fit of \textit{C(T)/T} vs. $T^2$ below 10~K of the heat capacity data of LaAlGe (not shown) results in a very small electronic specific heat coefficient $\gamma$~=~0.93~mJ/(mol~K$^2$) and Debye temperature $\Theta_d$~=~436~K. The electronic specific heat coefficient of CeAlGe is much harder to obtain from the linear fit of the heat capacity data due to the magnetic order feature. As the best estimate of the electronic specific heat coefficient, we take the lowest value of the $C_p(T)/T$ above the magnetic order, which is 50~mJ/(mol~K$^2$). Low $\gamma$ values may be indicative of low carrier concentration, as is confirmed by the Hall measurement discussed below. The estimated magnetic entropy associated with the magnetic transition is about 0.75$R$ln2 indicating a doublet ground state.

For the conventional collinear AFM with single-ion anisotropy, the low-temperature electronic part of the $C_p(T)$ data can be presented as $C_m(T)\approx (1/T^2)e^{-\Delta/T}$, where $\Delta$ is the phenomenological gap. The fit of the low-temperature of $C_m(T)T^2$ vs $T^{-1}$ data is shown in the inset to Fig.~\ref{C}(a). The gap was estimated to be 9.11~$\pm$~0.09~K. 

Temperature-dependent heat capacity data for CeAlGe single crystal for the magnetic field applied along the \textit{c}-axis is shown in Fig.~\ref{C}(b). Once the magnetic field is increased the magnetic order is suppressed and, starting from 20~kOe, a broad maximum, a Schottky-type anomaly, is observed. This maximum shifts to higher temperature as the field is increased. Such behavior of the heat capacity data was also observed in CeGe$_{1.76}$\cite{Budko2014} and CeAl$_2$\cite{Bredl1978} and was associated with Zeeman splitting of a ground state doublet.

%----------------------------------------------------------------------------------------------------------
%----------------------------------------------------------------------------------------------------------
\subsection{Transport} 
Temperature-dependent resistivity data at $H$~=~0 for CeAlGe and LaAlGe are shown in Fig.~\ref{R}(a). The inset shows the enlarged low-temperature part of the data with the maximum representing the magnetic transition. The residual-resistivity-ratio of CeAlGe and LaAlGe single crystals is $\sim$~2, which can indicate either a disordered system or a system with low carrier concentration. We believe that the latter is the case here since the carrier concentration estimated from the Hall resistivity, shown in Fig.~\ref{R}(b), is 1.44~$\times$~10$^{20}$~cm$^{-3}$ (independent of temperature) which is 2 orders of magnitude smaller than that for copper, and comparable to that of graphite, arsenic and YbPtBi\cite{Hundley1997}, indicating that CeAlGe is a semimetal. The concentration of carriers, in this case holes because Hall effect is positive for CeAlGe, was estimated using one-band model with effective carrier concentration, $R_H=1/(e\,p_\textmd{\tiny eff})$. 

The Hall coefficient of CeAlGe is almost temperature independent up to 300~K. The inset to Fig.~\ref{R}(b) shows the features in the field-dependent Hall resistivity at 2 and 3~K associated with the magnetic order. 

\begin{figure}[t]
\centering
\includegraphics[width=1\linewidth]{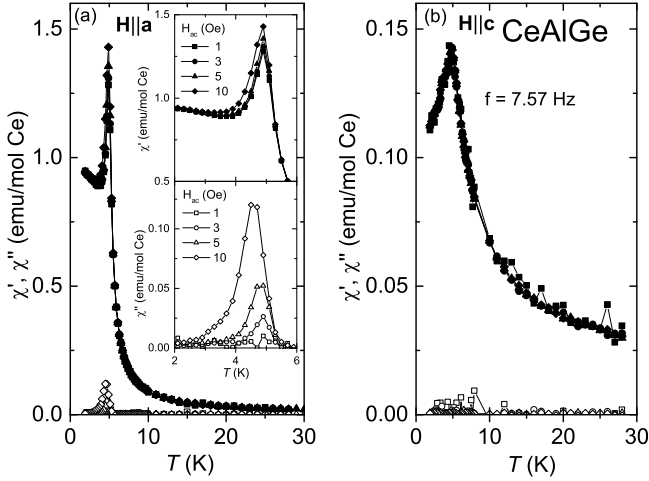}
\caption{\footnotesize (color online) Temperature-dependent $\chi'$ (closed symbols) and $\chi''$ (open symbols) for CeAlGe single crystal. (a) \textbf{H$_{ac}$}$\|$\textbf{a} and (b) \textbf{H$_{ac}$}$\|$\textbf{c} measured with varied $H_{ac}$.}
\label{AC}
\end{figure}

\begin{figure}[t]
\centering
\includegraphics[width=1\linewidth]{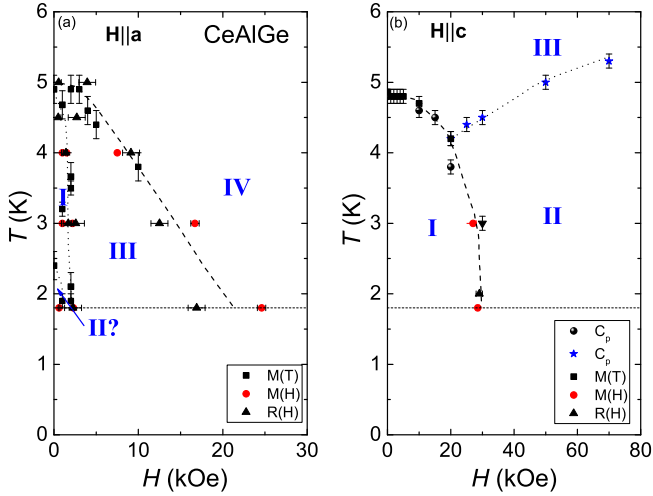}
\caption{\footnotesize (color online)  \textit{T}-\textit{H} phase diagrams for CeAlGe single crystals (a) \textbf{H}$\|\textbf{a}$ and (b) \textbf{H}$\|\textbf{c}$. Different phases are marked with roman numerals. Horizontal dashed lines indicate the temperature limit of our measurements. a) Phase III represents canted moment phase of the spin-flop transition. Phase IV represents a spin-flop to a partially saturated moment ($<$~1~$\mu_B$) phase.}
\label{PD}
\end{figure}

As is shown in Fig.~\ref{R}(c), the magnetoresistance of LaAlGe is positive reflecting that of a normal metal. There is no magnetic field effect on the resistivity of LaAlGe up to 5 K and 140~kOe. On the other hand, application of the magnetic field to CeAlGe has a profound effect, Fig.~\ref{R}(d). The magnetoresistance of CeAlGe is negative on the whole field range measured up to 12~K. The absolute value of the magnetoresistance decreases as the temperature is increased and it is almost zero at 20~K. The spin-flop and metamagnetic transitions in the $\rho(H)$ data collected at $T$~=~1.8~K (\textbf{H}$\|\textbf{a}$) and 2~K (\textbf{H}$\|\textbf{c}$) are consistent with the features associated with the spin-flop and metamagnetic transitions observed at the similar critical fields in the $M(H)$ data presented in Fig.~\ref{MH1}(c) and (d), respectively. 

\section{Discussion}

CeAlGe was previously reported to form in the tetragonal crystal structure with two different space groups: noncentrosymmetric I4$_1$md \cite{Zhao1990,Dhar1992,Dhar1996} and centrosymmetric I4$_1$/amd\cite{Flandorfer1998}. While the former structure will provide a potential route for the realization of nontrivial topology in this system, the latter structure will not. Thus, it is important to address the issue of the space group in this system especially since the easily accessible and widely used powder x-ray diffraction measurements are not able to distinguish between the two. With the help of single crystal x-ray diffraction, we were able to determine that CeAlGe single crystals most likely form in  the noncentrosymmetric I$4_1$md space group. 

We also mentioned above contradictory reports of both AFM and FM magnetic ground states in CeAlGe. For example, Ref.~\onlinecite{Flandorfer1998} suggests that CeAlGe forms in $\alpha$-ThSi$_2$-type structure (I4$_1$/amd space group) and the compound is suggested to be a soft ferromagnet with ordering temperature $T_C$~=~5.6~K, which is higher than the ordering temperature reported here, and the lattice parameters are smaller than what we observe in our work. This suggests Si sunstitution, since $T_C$ of CeAlSi is 9.6~K and the lattice parameters are smaller\cite{Bobev2005} than those of CeAlGe. The theoretical work also predicted CeAlGe to adopt a FM ground state.\cite{Chang2018} Our ZFC and FC magnetization data do not support FM order, but rather AFM order at $\sim$5~K. We did not observe hysteresis characteristic of a soft or hard FM in the \textit{M(H)} curves for both field orientations. The hysteresis in the \textit{M(H)} curve at 1.8~K for \textbf{H}$\|$\textbf{a}, Fig.~\ref{MH1}(c) is perhaps due to the reorientation of the domains as the system goes through the first-order spin-flop transition. We also did not observe anomalous Hall effect that is a signature of ferromagnets. The Arrot plot, $M^2$ versus \textit{H/M}, that provides an easy way of determining the presence of ferromagnetic order, does not hold. To further confirm the type of the magnetic order, we measured AC susceptibility, shown in Fig.~\ref{AC}. Just like DC temperature-dependent magnetization shown in Fig.~\ref{MH}, $\chi'$ (the real part of the AC susceptibility) and $\chi''$ (the imaginary part of the AC susceptibility) show large anisotropy, with $\chi'$ and $\chi''$ for magnetic field along the \textit{a}-axis being larger than those for the field along the \textit{c}-axis. $\chi'$ [\textbf{H}$\|$\textbf{a}, Fig.~\ref{AC}(a)] is slightly dependent on the AC drive field, while $\chi{''}$ shows much stronger AC-drive field dependent maximum. (Note that the absolute value of $\chi{''}$ is rather small.) These observations are consistent with the \textit{ferri}magnetic nature of the spin-flop transition.\cite{Balanda2013} On the contrary, the $\chi{'}$ for \textbf{H$_{ac}$}$\|$\textit{c} does not depend on $H_{ac}$ and $\chi{''}$ is almost zero, confirming AFM order.

The \textit{T-H} phase diagrams for \textbf{H}$\|$\textbf{a} and \textbf{H}$\|$\textbf{c} are shown in Figs.~\ref{PD} (a) and (b), respectively, and indicate that relatively small magnetic field on the order of 3 (\textbf{H}$\|$\textbf{a}) and 30~kOe (\textbf{H}$\|$\textbf{c}) is necessary to suppress the magnetic order. While the \textit{T-H} phase diagram for \textbf{H}$\|$\textbf{c} is relatively simple, the \textit{T-H} phase diagram for \textbf{H}$\|$\textbf{a} shows multiple phases due to the spin-flop transition (measurements below 1.8~K are necessary to confirm and delineate the phase II in more detail). If one naively assumes that the moment configuration in phase I is $\uparrow\downarrow\uparrow\downarrow$, then phase III represents canted moments $\uparrow\downarrow\nearrow\swarrow$ of the spin-flop transition and phase IV is then $\uparrow\downarrow\uparrow\uparrow$ according to our \textit{M(H)} data Fig. ~\ref{MH1}(c). Angular-dependent magnetization can shed more light on how the moment is directed/distributed in the tetragonal \textit{ab} plane of the noncentrosymmetric CeAlGe. Of course neutron measurements would be very helpful in assessing the magnetic structure and moment configuration of CeAlGe in zero and applied magnetic fields. It would also be interesting to further study quantum criticality for (\textbf{H}$\|$\textbf{c}), either with application of magnetic field or pressure.

\section{Conclusion}
In conclusion, we synthesized the first single crystals of CeAlGe and confirmed the noncentrosymmetric crystal structure required for type-II Weyl semimetal prediction. We determined that the compound assumes an AFM ground state below $\sim$~5~K. The magnetization is highly anisotropic, with \textit{a}-axis being an easy axis indicating that the magnetic moment lies in the tetragonal \textit{ab} plane. We determined, from the fit of the temperature-dependent heat capacity, the small gap in the magnon density of states to be 9.11~$\pm$~0.09~K. The small carrier density of 1.44~$\times$~10$^{20}$~cm$^{-3}$ indicates that CeAlGe is a semimetal.

%----------------------------------------------------------------------------------------------------------
%----------------------------------------------------------------------------------------------------------

\section{Acknowledgment} 

This work was supported by the Gordon and Betty Moore Foundation's EPiQS Initiative through grant no. GBMF4419.

\bibliography{papers}

\end{document}